# Generalized two-temperature model for coupled phonons

Meng An[1,2,#], Qichen Song[1,2,#,†], Xiaoxiang Yu[1,2], Han Meng[1,2], Dengke Ma[1,2], Ruiyang Li[1,2], Zelin Jin[1,2], Baoling Huang[3,4], and Nuo Yang[1,2*]

[1]State Key Laboratory of Coal Combustion, Huazhong University of Science and Technology, Wuhan 430074, People's Republic of China

[2]Nano Interface Center for Energy (NICE), School of Energy and Power Engineering, Huazhong University of Science and Technology (HUST), Wuhan 430074, People's Republic of China

[3]Department of Mechanical and Aerospace Engineering, The Hong Kong University of Science and Technology, Clear Water Bay, Kowloon, Hong Kong

[4]The Hong Kong University of Science and Technology Shenzhen Research Institute, Shenzhen, 518057, China

# M.A. and Q.S. contributed equally to this work.

† Current address: Department of Mechanical Engineering, Massachusetts Institute of Technology, 77 Massachusetts Avenue, Cambridge, MA 02139, USA.

* To whom correspondence should be addressed. E-mail: (N.Y.) nuo@hust.edu.cn

**Abstract**

The design of graphene-based composite with high thermal conductivity requires a comprehensive understanding of phonon coupling in graphene. We extended the two-temperature model to coupled groups of phonon. The study give new physical quantities, the phonon-phonon coupling factor and length, to characterize the couplings quantitatively. Besides, our proposed coupling length has an obvious dependence on system size. Our studies can not only observe the nonequilibrium between different groups of phonon, but explain theoretically the thermal resistance inside graphene.

## Introduction

Owing to the superior thermal conductivity 3000-5000 W/m-K of graphene[1], the graphene-based composites[2-5] have been widely applied in thermal managements of advanced electronics[6], optoelectronics, the photovoltaic solar cell[7] and Li-ion battery[4]. For example, it is found that the thermal conductivity of graphene-based composites at relatively low filler contents can reach up 5.1 W/m-K[2]. Besides, it has been demonstrated that thermal conductivity of graphene-based composite can be further enhanced by changing the layer number of graphene sheets or flakes and the orientation[4,8,9]. To enable rational design of graphene-based composites with higher thermal conductivity, a fundamental and comprehensive understanding of thermal transport in such materials is essential.

Extensive studies have concluded that two main thermal resistances play central roles in determining the thermal conductivity of graphene-based composites. The first one is the high interfacial thermal resistance between graphene and matrix materials[2,10]. Recent investigation showed that the interfacial thermal resistance could be overcome by alignment arrangement of graphene in composites[6]. The second one is the phonon coupling thermal resistance between low-frequency out-of-plane (OP) phonons and high-frequency in-plane (IP) phonons[11], which is not well studied and responsible for the thermal resistance inside graphene. The heat energy is mostly transferred from matrix materials to OP phonons of graphene due to the strong coupling between the matrix materials and the low-frequency OP phonons (shown in Fig. S1). In graphene-based composite samples, the size of graphene flakes/particles is nanoscale[2,6]. The studies[12,13] indicated that, in graphene nanoflakes/particles, the OP phonons contribute less to the thermal conductivity than IP phonons because the OP phonons have longer mean free path and wave length than IP phonons and are suppressed more by the size confinement. So, there is an indirect heat transport inside graphene nanoflakes/particles: the heat will be transferred through the coupling between OP and in-plane (IP) phonons. However, the coupling between OP and IP is relatively weak[14,15]. As a result, that leads to a large thermal resistance inside graphene. In this work, we mainly focus on the phonon mode coupling resistance between IP and OP, which determines the heat flux inside graphene.

The two-temperature model (TTM) has been successfully applied to investigate the coupling between different energy carriers in thermal nonequilibrium[16,17]. In TTM, different energy carriers are considered as two interacting subsystems, such as the electron-phonon[18,19] and phonon-magnon[16,20]. Liao et al. have generalized TTM for the coupled phonon-magnon diffusion and predicted a magnon cooling effect[16]. Besides, some reports show TTM is implemented with molecular dynamics which is used to study electron-phonon coupled in metal/semiconductor systems[19,21,22].

Here, we extended TTM to investigate the coupling between different phonon groups. This method is demonstrated on graphene, a representative two-dimensional material with the highest known thermal conductivity[23,24]. We separate phonons in graphene into two groups, namely in-plane (IP) and out-of-plane (OP) phonon group. Based on the temperature profiles of different phonon groups, we calculated the ph-ph coupling factor $G_{io}$, comparable with strength of electron-phonon coupling[19] and found that $G_{io}$ is not sensitive to system size. We successfully demonstrated the poor coupling between IP and OP phonon groups in graphene[25-28].

**Theoretical analysis**

In this letter, TTM is used to investigate the ph-ph coupling between OP and IP phonon groups. As shown in the configuration (Fig. 1), the entire system is divided into two regions, namely, the left region ($l$) (where atoms have both OP and IP vibrations) and the right region (r) (only IP vibration). $T_H$ ($T_C$) is the prescribed temperature of hot (cold) bath at the left (right) end. As observed in the left region, the temperature difference between OP and IP phonon groups increases and reaches a maximum value, $\theta_{max}$, at the interface. This phenomenon results from the poor coupling between OP and IP phonon groups, which is a necessary condition for TTM.

The Boltzmann transport equation (BTE) for IP and OP phonons in graphene that determines the phonon distribution function can be written separately as:

$$\frac{\partial f_i}{\partial t} + \mathbf{F} \cdot \nabla_{\mathbf{p}} f_i + \mathbf{v} \cdot \nabla_{\mathbf{r}} f_i = \left( \frac{\partial f_i}{\partial t} \right)_c \tag{1a}$$

$$\frac{\partial f_o}{\partial t} + \mathbf{F} \cdot \nabla_{\mathbf{p}} f_o + \mathbf{v} \cdot \nabla_{\mathbf{r}} f_o = \left( \frac{\partial f_o}{\partial t} \right)_c \tag{1b}$$

where the subscripts $o$ and $i$ denote OP and IP phonons, respectively. When the system reaches steady state without applying external force, the Eq. (1a) and Eq. (1b) can be simplified as:

$$\mathbf{v} \cdot \nabla_{\mathbf{r}} f_i = \left( \frac{\partial f_i}{\partial t} \right)_c \tag{2a}$$

$$\mathbf{v} \cdot \nabla_{\mathbf{r}} f_o = \left( \frac{\partial f_o}{\partial t} \right)_c \tag{2b}$$

where the right-hand sides are the collision term, which includes three kinds of phonon scattering processes: IP-IP scatterings, OP-OP scatterings and scatterings between IP and OP phonons.

Previous studies[29,30] have demonstrated that the reflection symmetry in graphene forbids odd number of OP phonons (ZA and ZO) in a three-phonon scattering process, which leads to the relatively weak coupling between OP and IP phonons. Moreover, we also observed the weak couplings by MD simulation results (shown in SI and Fig. S4). The weak coupling has a negligible effect on the distribution and phase space of phonons. IP and OP phonons will be treated as two subsystems. Then, the collision term can be written as:

$$\mathbf{v} \cdot \nabla_{\mathbf{r}} f_i = \left( \frac{\partial f_i}{\partial t} \right)_{io} + \left( \frac{\partial f_i}{\partial t} \right)_{ii} \tag{3a}$$

$$\mathbf{v} \cdot \nabla_{\mathbf{r}} f_o = \left( \frac{\partial f_o}{\partial t} \right)_{io} + \left( \frac{\partial f_o}{\partial t} \right)_{oo} \tag{3b}$$

where the subscript *ii*, *oo*, *io* denote IP-IP scatterings, OP-OP scatterings and IP-OP scatterings, respectively. When the relaxation time approximation (RTA) is adopted for IP-IP scatterings, OP-OP scatterings, the Eq. (3a) and Eq. (3b) can write:

$$\mathbf{v} \cdot \nabla_{\mathbf{r}} f_i = \left( \frac{\partial f_i}{\partial t} \right)_{io} + \left( \frac{f_i - f_{i,0}}{\tau_{ii}} \right)_{ii} \tag{4a}$$

$$\mathbf{v} \cdot \nabla_{\mathbf{r}} f_o = \left( \frac{\partial f_o}{\partial t} \right)_{io} + \left( \frac{f_o - f_{o,0}}{\tau_{oo}} \right)_{oo} \tag{4b}$$

where $f_{i,0}$ and $f_{o,0}$ is the IP and OP phonon distribution function at equilibrium (the Bose-Einstein distribution $f = \left( \exp\left( h\omega / k_B T \right) - 1 \right)^{-1}$, respectively. $\tau_{ii}$ and $\tau_{oo}$ are the relaxation time for IP and OP phonons, respectively. After multiplying Eq. (4a) and Eq. (4b) by ℏω, and integrating over all wavevector $\mathbf{q}$, the last terms on the right-hand sides of Eq. (4a) and Eq.

(4b) drop out because they are odd function with respect to all wavevector $\mathbf{q}$ [31]. And they are described as:

$$\nabla_{\mathbf{r}} \cdot \sum_{\mathbf{q}} \hbar \omega_{\mathbf{q}} \mathbf{v} f_i = \sum_{\mathbf{q}} \hbar \omega_{\mathbf{q}} \left( \frac{\partial f_i}{\partial t} \right)_{io} \tag{5a}$$

$$\nabla_{\mathbf{r}} \cdot \sum_{\mathbf{q}} \hbar \omega_{\mathbf{q}} \mathbf{v} f_o = \sum_{\mathbf{q}} \hbar \omega_{\mathbf{q}} \left( \frac{\partial f_o}{\partial t} \right)_{io} \tag{5b}$$

Denoting $\mathbf{J}_{i/o} = \sum_{\mathbf{q}} \hbar \omega_{\mathbf{q}} \mathbf{v} f_{i/o}$ and $\partial E_{i/o} / \partial t = \sum_{\mathbf{q}} \hbar \omega_{\mathbf{q}} \left( \partial f_{i/o} / \partial t \right)_{io}$, they write:

$$\nabla \cdot \mathbf{J}_i = -\frac{\partial E_i}{\partial t} \tag{6a}$$

$$\nabla \cdot \mathbf{J}_o = -\frac{\partial E_o}{\partial t} \tag{6b}$$

The scatterings between IP-OP are responsible for the local energy exchange between IP and OP. We use $1 / \tau_{io}$ to describe the scattering rate for the coupling between IP and OP. Then the rate of energy transfer between IP and OP is then described by[32]:

$$\frac{\Delta E_i}{\Delta t} = \frac{c_i \left( T_{inter} - T_i \right)}{\tau_{io}} \tag{7a}$$

$$\frac{\Delta E_o}{\Delta t} = \frac{c_o \left( T_{inter} - T_o \right)}{\tau_{io}} \tag{7b}$$

where the intermediate temperature that both channels approach is $T_{inter} = \dfrac{c_i T_i + c_o T_o}{c_i + c_o}$. Therefore, the rate of energy transfer is written as:

$$\frac{\Delta E_i}{\Delta t} = \frac{c_i c_o \left( T_o - T_i \right)}{\tau_{io} \left( c_i + c_o \right)} = G_{io} \left( T_o - T_i \right) \tag{8a}$$

$$\frac{\Delta E_o}{\Delta t} = \frac{c_i c_o (T_i - T_o)}{\tau_{io}(c_i + c_o)} = G_{io}(T_i - T_o) \tag{8b}$$

$$G_{io} = \frac{c_i c_o}{\tau_{io}(c_i + c_o)} \tag{8c}$$

where $G_{io}$, is defined as the coupling factor. Then, with the heat diffusion equation, the Eq. (7a) and Eq. (7b) can be written as:

$$\nabla \mathbf{q}_i = G_{io}(T_o - T_i) \tag{9a}$$

$$\nabla \mathbf{q}_o = G_{io}(T_i - T_o) \tag{9b}$$

We then apply the Fourier Law $\mathbf{q} = -\kappa \nabla T$ to rewrite the Eq. (9a) and Eq. (9b):

$$-\kappa_i \nabla^2 T_i = G_{io}(T_o - T_i) \tag{10a}$$

$$-\kappa_o \nabla^2 T_o = G_{io}(T_i - T_o) \tag{10b}$$

Considering the one-dimensional temperature gradient in graphene, the two-dimensional heat transfer problem through IP, OP and the coupling between OP and IP can be well described by a one-dimensional (1D) model. The governing equations for coupled phonon transport states:

$$\kappa_o \frac{d^2 T_o}{dx^2} = G_{io}(T_o - T_i) \tag{11a}$$

$$\text{-} \kappa_i \frac{d^2 T_i}{dx^2} = G_{io}(T_o - T_i) \tag{11b}$$

Subtracting Eq. (11a) from Eq. (11b), we can obtain $d^2\theta / dx^2 - \gamma^2 \theta = 0$, where $\theta = T_o - T_i$ and $\gamma = \sqrt{G_{io}(1/\kappa_i + 1/\kappa_o)}$. Based on the boundary condition $\theta\big|_{x \to -\infty} = 0$ and $\theta\big|_{x=0} = \theta_{\max}$, the temperature difference profile is written as $\theta = \theta_{\max} \exp(\gamma x)$.

The validity of TTM requires that the coupling between two carriers is so weak that the coupling has a negligible effect on the distribution and phase space of each carrier. As far as we know, TTM has been successfully used to describe the coupling of electron-phonon[18,19], and phonon-magnon[16,20]. For the coupling between IP phonons and OP phonons, both them are bosons, which is similar the phonon-magnon coupling. Both the experimental and simulation work[14,15] have implied the coupling of IP-OP is much weaker than that of either IP-IP or OP-OP. In addition, this weak coupling is confirmed by our MD results (shown in Fig. S2). It is deemed that the IP-OP coupling has a negligible effect on the distribution and phase space of either IP or OP phonon group. Therefore, the IP-OP coupling satisfies the requirement of TTM. We have computed more simulation results (Fig. S6) to show the strong coupling between two in-plane directions (x and y). Because TTM is only applicable to the weak coupling between different energy carriers, the coupling factor between two IP directions should not be resolved using TTM.

Here, we propose the ph-ph coupling length, $l_c$, shown in Fig. 1, to quantitatively characterize the coupling between them. Its physical meaning is the distance required to equilibrate OP and IP phonon groups when the temperature difference exists. Specifically, we define such a characteristic length as the distance between the position of $\theta_{max}$ and $5\%\theta_{max}$:

$$l_c = \frac{-\ln(5\%)}{\gamma} \approx \frac{3}{\gamma} = \frac{3}{\sqrt{G_{io}\left(\dfrac{1}{\kappa_i} + \dfrac{1}{\kappa_o}\right)}} \tag{12}$$

**Numerical Simulations**

The ph-ph coupling length of graphene is numerical calculated by means of TTM-MD, and the detailed non-equilibrium MD simulation setup is shown in Fig. 2(a). The heat source with a higher temperature $T_H$=310K is applied to the atoms in red region and the heat sink with a lower temperature $T_C$=290K is applied to atoms in blue region where only IP phonons exist. All of our simulations are performed by large-scale atomic/molecular massively parallel simulator (LAMMPS) packages[33]. The fixed (periodic) boundary conditions are used along the x (y) direction. The optimized Tersoff potential[34] is applied to describe interatomic interactions, which has successfully reproduced the thermal transport properties of graphene[35,36]. The detailed parameters of optimized Tersoff potential are shown in Supporting Information (SI). The velocity Verlet algorithm is adopted to integrate the discrete differential equations of motions. The time step is set as 0.5fs. We relaxed the graphene structure in the isothermal-isobaric (NPT) ensemble. After NPT relaxation, simulations are performed for 2ns to reach a steady state. After that, a time average of the temperature and heat current is performed for 10ns.

The thermal conductivity is calculated based on the Fourier's Law[37] $\kappa = -J/A \cdot \nabla T$, where J denotes the heat current transported from the hot bath to cold bath, A is the cross section area and $\nabla T$ is the temperature gradient along the x direction. The temperature of different phonon groups is computed by $T_p(x) = \left\langle \sum_{i=1}^{N} m_i \mathbf{v}_{i,p} \mathbf{v}_{i,p} \right\rangle / N k_B$, where $\mathbf{v}_{i,p}$ is the velocity vector of phonon vibrating along $p$ direction, i.e. IP or OP, of atom $i$. $N$ and $k_B$ are the number of atoms in the simulation cell and the Boltzmann constant, respectively. Besides, it is calculated that the thermal conductivity of IP (OP) including only the atomic vibration along in-plane (out-of-plane) direction by freezing the other direction. The simulations detail is shown in Supplementary Information (Fig. S3).

## Results and Discussion

In our simulations, the width of simulation cell is set as 5.2 nm. The lattice constant (a) and thickness (d) of graphene are 0.143 nm and 0.335 nm, respectively. The size effect could arise if the width is not sufficiently large[38,39]. The thermal conductivity reaches a saturated value when the width is larger than 5.2 nm which is consistent with our previous works[39,40]. Besides, to examine the accuracy of MD results we calculated thermal conductivity ($\kappa$) of suspended single-layer graphene with a length (L) as 33 nm. The calculated $\kappa$ is $1082 \pm 103$ W/m-K, which is consistent with previous studies[35,36].

The temperature profile of TTM-MD is presented in Fig. 2(b), where the system length along x direction is 101 nm. In the left region, there are obvious differences between the temperature curves of IP and OP phonons. It is found that the maximum temperature difference ($\theta_{max}$) is 5.1 K at the interface. In order to extract the ph-ph coupling length ($l_c$), the profile of $\theta$ in Fig. 2(c) is fitted exponentially, based on theoretical solution of Eq. (11a and 11b). Then, $l_c$ can be obtained based on its definition as the distance between the position of $\theta_{max}$ and 5%$\theta_{max}$. If the system length is not large enough to equilibrate different phonon groups, the coupling length would be larger than system length. That is IP and OP subsystems are under nonequilibrium state. We calculated four simulation cells with different length, and $l_c$ are 60 nm, 64 nm, 70 nm and 77 nm corresponding to length is 33 nm, 60 nm, 67 nm, and 101 nm, respectively. The profiles of temperature difference with L=33nm, 67nm, 70nm is shown in Supplementary Information (Fig. S4).

To evaluate the ph-ph coupling factor ($G_{io}$) based on Eq. (12), we calculated the thermal conductivity of IP/OP phonon groups ($\kappa_i/\kappa_o$) including only the atomic vibration along in-plane (out-of-plane) direction (shown in Fig. 3). It is also shown that the summation of $\kappa_i$ and $\kappa_o$ approximately equals to the thermal conductivity of pristine graphene (blue open circles), instead of an obvious reduction by scatterings. This phenomenon in graphene arises from the poor ph-ph coupling between OP and IP[26,28,41] and the decoupled phonon scattering interaction between OP and IP of freezing method[28]. Moreover, it is noted that our MD results on graphene with nano-confinement shows (in Fig. 3) the relative contribution of OP phonons is around 25%, which agrees with the results from other groups by theoretical prediction[42] and MD results[13,43,44]. On the other side, Lindsay *et al.*[30] found that, in an infinite graphene sheet, the OP phonons in graphene dominate the thermal transport because the reflection symmetry in two-dimensional graphene significantly restricts the phase space for phonon-phonon scattering of OP phonons[45]. This discrepancy may come from three possible reasons. The first is the presence of normal scatterings which is excluded in the iterative three-phonon calculation. The second reason may come from the essential difference between three-phonon calculation and MD simulation. In anharmonic lattice dynamics calculation, the graphene sheet remains a perfect plane with the reflection symmetry perfectly preserved, while in MD simulations the atoms are not at their equilibrium positions and the graphene is not a plane. As a result, the reflection symmetry may not be well presented in MD simulations. Thirdly, another possible reason is the size effect. That is, Lindsay et.al studied bulk graphene[45], while our MD results and other groups calculated the thermal conductivity of finite-sized nanostructured graphene.

To show that our results are robust, we also have used the AIREBO potential[46] to calculate the thermal conductivity of different phonon groups in graphene (as shown in Fig. S5). It is found that the relative contribution of OP phonons is smaller than IP phonons for both two potentials.

In low dimensional nanostructures, the thermal conductivity becomes dependent on the characteristic length of system due to the length is comparable to phonon mean free path (MFP)[47]. Recent studies[9,23,48-50] showed that the thermal conductivity of graphene increases logarithmically (~log$L$) with the size of sample, which is taken to fit MD results shown in Fig. 3.

With the $\kappa_i$, $\kappa_o$ and $l_c$ for several systems with different lengths, we can calculate the ph-ph coupling factor $G_{io}$ based on Eq. (12) (shown in Fig. 4(a)). To the best of our knowledge, it is the first time to extract the ph-ph coupling factor. It is found that $G_{io}$ in graphene is not sensitive to system length, and the average value is $4.26\times10^{17}$ W/m$^3$-K, which is comparable with the electron-phonon coupling factor ($G_{ep}$) in metals ranging from $5.5\times10^{16}$W/m$^3$-K to $2.6\times10^{17}$W/m$^3$-K[19]. To verify our results, the coupling factor between IP and OP phonon groups is also calculated from anharmonic lattice dynamics and first-principles (calculation details in SI). The $G_{io}$ is calculated as $1.07\times10^{17}$ W/m$^3$-K , which is the same order of magnitude as the value from MD simulation, $4.26\times10^{17}$ W/m$^3$-K .The discrepancy may come from the fact that the reflection symmetry in graphene forbids odd number of OP modes (ZA and ZO) in a three-phonon scattering process[29,30]. As a result, the interaction between IP and OP phonon can be weak. In first-principle calculation, this reflection is forced to be preserved, yet in MD simulation and in reality, the existence of boundary can break the reflection symmetry and induces more scattering that is involved with both IP and OP phonons. Besides, the first-principles calculation used here fails to capture higher-order phonon scattering, which is naturally embedded in MD simulation. This may also lead to the underestimated coupling factor from first-principle.

As shown in Fig. 4(b), we obtained the size dependence of ph-ph coupling length of graphene based on the size dependence of the thermal conductivity. It is clearly observed that the ph-ph coupling length, $l_c$, diverges when $\kappa \sim log$L[23]. Combining with the definition of $l_c$ in Eq. (12), it is easy to understand that the size dependence of thermal conductivity contributes to the size effect of $l_c$.

The ph-ph coupling factor and length may provide us a new way in understanding heat transfer. For example, the weak coupling means there are less scatterings between IP and OP phonons, which attribute to the ultra-high thermal conductivity of suspended graphene sheet. The coupling factor and length provide a perspective in explanation quantitatively. As we know, the thermal performance of graphene-based composites is determined from the interfacial thermal resistance (named as outer resistance)[51,52]. Besides, based on our finding above, the longer ph-ph coupling length means it is difficult to reach the equilibrium state between IP and OP phonons, which corresponds to a larger inside thermal resistance. Both the outer and inside resistances contribute to the unsatisfying thermal performance of graphene-based composites.

**Conclusion**

In summary, we have extended the TTM to investigate the coupled phonons implemented with molecular dynamics. The merit of this work lies in the fact that we applied the techniques widely used in the field of electron-phonon interactions to investigate the ph-ph interactions, utilized the analogy between field-driven electron and phonon, and combined IP and OP phonon groups in a smooth way. The magnitude of ph-ph coupling factor has been estimated in graphene and is independent of system size. Besides, we proposed the ph-ph coupling length associated with the ph-ph coupling strength. Our studies can not only observe the nonequilibrium between different groups of phonon, but explain theoretically the thermal resistance inside graphene.

**Acknowledgements**

This project is supported by the National Natural Science Foundation of China (Grant 51576076). We are grateful to Jingtao Lü, Gang Zhang, Lifa Zhang, Jinlong Ma and Shiqian Hu for useful discussions. The authors thank the National Supercomputing Center in Tianjin (NSCC-TJ) and China Scientific Computing Grid (ScGrid) for providing assistance in computations.

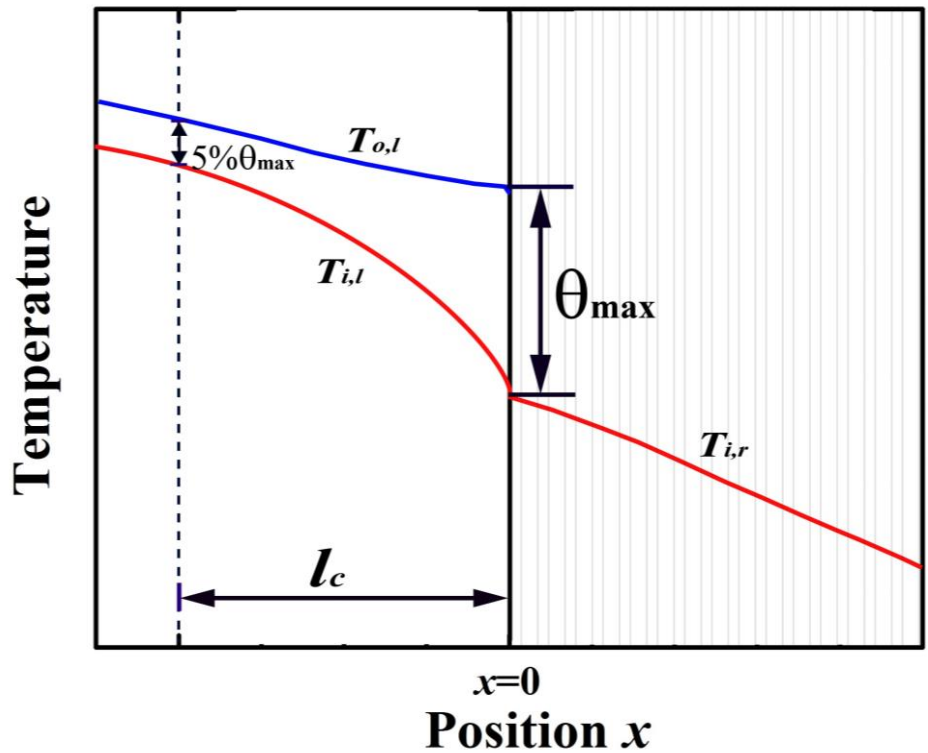

Figure 1 (Color on-line) Representative temperature profile in TTM for the coupled phonons. The system consists of two regions, namely, the left region (*l*) (where atoms have both OP and IP vibrations) and the right region (r) (only IP vibration). $T_{o,l}$, $T_{i,l}$ are temperatures for OP and IP phonon group in the left region. $T_{i,r}$ is the temperature of IP phonon group in the right region. $\theta_{max}$ is the maximum temperature difference between OP and IP phonon group. $l_c$ is the ph-ph coupling length, which is defined as the distance between the position of $\theta_{max}$ and $5\%\theta_{max}$.

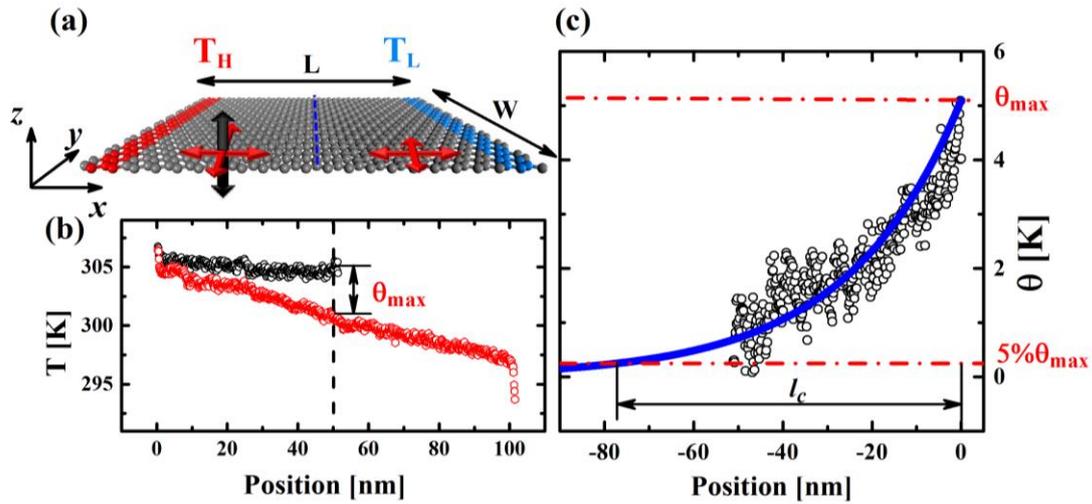

Figure 2 (Color on-line) (a) Schematic illustration of molecular dynamics simulation setup. The black (red) arrow denotes the vibration along the out-of-plane (in-plane) direction. The blue dot line is the interface of different simulation regions. The atoms in the left region vibrate freely in three directions while those in the right region could only vibrate along in-plane direction. The fixed (periodic) boundary condition is applied to along x (y) direction. Hot bath (red atoms) $T_H$ and cold bath (blue atoms) $T_C$ are applied. The temperatures of two heat baths are set as $T_H$=310K, $T_C$=290K, respectively. (b) The temperature profile of different phonon groups. (c) The temperature difference distribution between OP and IP phonon group when the system length is set as L=101nm. The fitting line is based on $\theta = \theta_{max} \exp(\gamma x)$. The red dot lines denote the maximum temperature difference and its five percent, respectively.

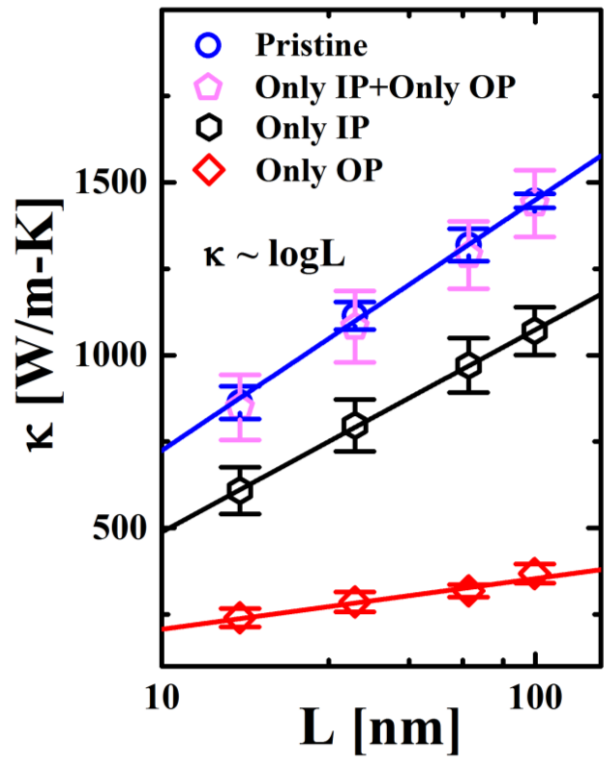

Figure 3. (Color on-line) The length dependence of thermal conductivity with different phonon groups in graphene calculated from NEMD. Pristine graphene (Pristine), atoms vibrated along two in-plane directions only (Only IP), atoms vibrated along our-of-plane direction only (only OP), and the summation of the values of only IP and only OP. The system length ranges from 16 nm to 100 nm. The fitting lines are based on κ ~ logL.

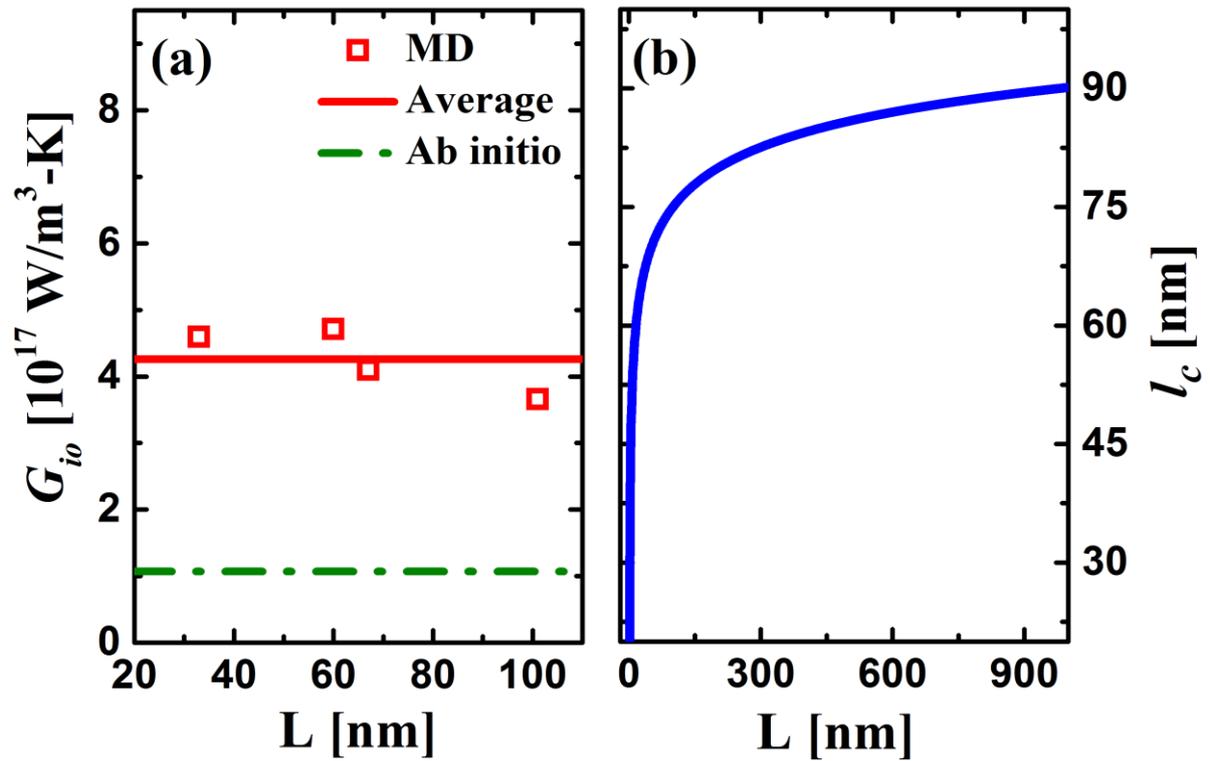

Figure 4. (Color on-line) (a) The ph-ph coupling factor versus the system lengths calculated by MD and ab initio methods. (b) The length dependence of the ph-ph coupling length from MD simulation, where the thermal conductivity of graphene follows $\kappa \sim \log L$.

# References

[1]A.A. Balandin, S. Ghosh, W. Bao, I. Calizo, D. Teweldebrhan, F. Miao, and C.N. Lau, Nano Lett. **8**, 902 (2008).

[2]K.M. Shahil and A.A. Balandin, Nano Lett. **12**, 861 (2012).

[3]J.D. Renteria, D.L. Nika, and A.A. Balandin, Appl. Sci. **4**, 525 (2014).

[4]P. Goli, S. Legedza, A. Dhar, R. Salgado, J. Renteria, and A.A. Balandin, J. Power Sources **248**, 37 (2014).

[5]D.L. Nika and A.A. Balandin, Rep. Prog. Phys. **80**, 036502 (2017).

[6]J. Renteria, S. Legedza, R. Salgado, M. Balandin, S. Ramirez, M. Saadah, F. Kargar, and A. Balandin, Materials & Design **88**, 214 (2015).

[7]G. Notton, C. Cristofari, M. Mattei, and P. Poggi, Appl. Therm. Eng. **25**, 2854 (2005).

[8]V. Goyal and A.A. Balandin, Appl. Phys. Lett. **100**, 073113 (2012).

[9]A.A. Balandin, Nat. Mater. **10**, 569 (2011).

[10]K.M.F. Shahil and A.A. Balandin, Solid State Commun. **152**, 1331 (2012).

[11]X. Wu and T. Luo, J. Appl. Phys. **115**, 014901 (2014).

[12]Z. Wei, J. Yang, K. Bi, and Y. Chen, J. Appl. Phys. **116**, 153503 (2014).

[13]B. Qiu and X. Ruan, Appl. Phys. Lett. **100**, 193101 (2012).

[14]S. Sullivan, A. Vallabhaneni, I. Kholmanov, X. Ruan, J. Murthy, and L. Shi, Nano Lett. **17**, 2049 (2017).

[15]A.K. Vallabhaneni, D. Singh, H. Bao, J. Murthy, and X. Ruan, Phys. Rev. B **93**, 125432 (2016).

[16]B. Liao, J. Zhou, and G. Chen, Phys. Rev. Lett. **113**, 025902 (2014).

[17]A. Majumdar and P. Reddy, Appl. Phys. Lett. **84**, 4768 (2004).

[18]G. Chen, J. Appl. Phys. **97**, 083707 (2005).

[19]Y. Wang, X. Ruan, and A.K. Roy, Phys. Rev. B **85**, 205311 (2012).

[20]K. An, K.S. Olsson, A. Weathers, S. Sullivan, X. Chen, X. Li, L.G. Marshall, X. Ma, N. Klimovich, and J. Zhou, Phys. Rev. Lett. **117**, 107202 (2016).

[21]Z.X. Lu, Y. Wang, and X.L. Ruan, Phys. Rev. B **93**, 064302 (2016).

[22]N. Yang, T. Luo, K. Esfarjani, A. Henry, Z. Tian, J. Shiomi, Y. Chalopin, B. Li, and G. Chen, J. Comput. Theor. Nanos. **12**, 168 (2015).

[23]X. Xu, L.F. Pereira, Y. Wang, J. Wu, K. Zhang, X. Zhao, S. Bae, C. Tinh Bui, R. Xie, J.T. Thong, B.H. Hong, K.P. Loh, D. Donadio, B. Li, and B. Ozyilmaz, Nat. Commun. **5**, 3689 (2014).

[24]A.A. Balandin, S. Ghosh, W. Bao, I. Calizo, D. Teweldebrhan, F. Miao, and C.N. Lau, Nano Lett. **8**, 902 (2008).

[25]J. Zhang, Y. Wang, and X. Wang, Nanoscale **5**, 11598 (2013).

[26]J. Zhang, X. Wang, and H. Xie, Phys. Lett. A **377**, 2970 (2013).

[27]J. Zhang, X. Huang, Y. Yue, J. Wang, and X. Wang, Phys. Rev. B **84**, 235416 (2011).

[28]H. Zhang, G. Lee, and K. Cho, Phys. Rev. B **84**, 115460 (2011).

[29]X. Gu, Y. Wei, X. Yin, B. Li, and R. Yang, arXiv preprint arXiv:1705.06156 (2017).

[30]L. Lindsay, D.A. Broido, and N. Mingo, Phys. Rev. B **82**, 115427 (2010).

[31]G. Chen, *Nanoscale energy transport and conversion: a parallel treatment of electrons, molecules, phonons, and photons*. (Oxford University Press, 2005).

[32]D. Sanders and D. Walton, Phys. Rev. B **15**, 1489 (1977).

[33]S. Plimpton, J. Comput. Phys. **117**, 1 (1995).

[34]L. Lindsay and D.A. Broido, Phys. Rev. B **81**, 205441 (2010).

[35]L. Yang, J. Chen, N. Yang, and B. Li, Int. J. Heat Mass Transfer **91**, 428 (2015).

[36]J. Chen, G. Zhang, and B. Li, Nanoscale **5**, 532 (2013).

[37]N. Yang, G. Zhang, and B. Li, Nano Lett. **8**, 276 (2008).

[38]P.K. Schelling, S.R. Phillpot, and P. Keblinski, Phys. Rev. B **65**, 144306 (2002).

[39]H. Shiqian, A. Meng, Y. Nuo, and L. Baowen, Nanotechnology **27**, 265702 (2016).

[40]Q. Song, M. An, X. Chen, Z. Peng, J. Zang, and N. Yang, Nanoscale **8**, 14943 (2016).

[41]J. Zhang, X. Wang, and H. Xie, Phys. Lett. A **377**, 721 (2013).

[42]D.L. Nika, S. Ghosh, E.P. Pokatilov, and A.A. Balandin, Appl. Phys. Lett. **94**, 203103 (2009).

[43]T. Feng, X. Ruan, Z. Ye, and B. Cao, Phys. Rev. B **91**, 224301 (2015).

[44]J.-H. Zou and B.-Y. Cao, Appl. Phys. Lett. **110**, 103106 (2017).

[45]L. Lindsay, W. Li, J. Carrete, N. Mingo, D.A. Broido, and T.L. Reinecke, Phys. Rev. B **89**, 155426 (2014).

[46]S.J. Stuart, A.B. Tutein, and J.A. Harrison, J. Chem. Phys. **112**, 6472 (2000).

[47]N. Yang, X. Xu, G. Zhang, and B. Li, AIP Advances **2**, 041410 (2012).

[48]S. Ghosh, W. Bao, D.L. Nika, S. Subrina, E.P. Pokatilov, C.N. Lau, and A.A. Balandin, Nat. Mater. **9**, 555 (2010).

[49]D.L. Nika, A.S. Askerov, and A.A. Balandin, Nano Lett. **12**, 3238 (2012).

[50]D.L. Nika and A.A. Balandin, J. Phys.: Condens. Matter **24**, 233203 (2012).

[51]Y. Liu, J. Huang, B. Yang, B.G. Sumpter, and R. Qiao, Carbon **75**, 169 (2014).

[52]L. Hu, T. Desai, and P. Keblinski, Phys. Rev. B **83**, 195423 (2011).